\magnification=1200

\font\bigbf=cmbx9  scaled\magstep3 \vskip 0.2in \centerline{\bigbf
A method of calculation for the  } \vskip 0.1in
\centerline{\bigbf determinant of the Potts model transfer matrix
}

\vskip 0.4in \font\bigtenrm=cmr10 scaled\magstep1
\centerline{\bigtenrm  B. Mirza and M.R. Bakhtiari} \vskip 0.2in

\centerline{\sl Department of Physics, Isfahan University of
Technology, Isfahan 84154, Iran } \vskip 0.1in

\centerline{\sl E-mail: b.mirza@cc.iut.ac.ir}

\vskip 0.2in \centerline{\bf ABSTRACT} \vskip 0.1in By using a
 decomposition of the transfer matrix of the two dimensional
 $q$-state Potts Model to $V^{\prime}_1$ and $V_2$ its determinant is calculated. Our
 result is a proof for a conjectured
 formula by Chang and Shrock in [14].



\vfill\eject

\vskip 1in \centerline{I. \bf   Introduction } \vskip 0.1in

The two dimensional $q$-state Potts models [1,2] for various $q$
have been of interest as examples of different universality
classes for phase transitions and, for $q=3,4$ as models for the
adsorption of gases on certain substrates [3,4,5]. For  $q \geq $3
the free energy has never been calculated in closed form for
arbitrary temperature. It is thus of continuing value to obtain
further information about the two dimensional Potts model. Some
exact results have been established for the model: from a duality
relation, the critical point has been identified [1]. The free
energy and latent heat [6,7,8], and magnetization [9] have been
calculated exactly by Baxter at this critical point, establishing
that the model has a continuous, second order transition for $q
\leq 4$  and a first order transition for $q \geq 5$ . Baxter has
also shown that although the  $q = 3$ model has no phase with
antiferromagnetic long-range order at any finite temperature there
is an antiferromagnetic critical point at $ T= 0$ [9]. The values
of the critical exponents (for the range of $q$ where the
transition is continuous) have been determined [10,11,12]. Further
insight into the critical behaviour was gained using the methods
of conformal field theory [13]. In this paper a proof for a
conjectured formula by Chang and  Shrock [14] for determinants of
the transfer matrices of the $q$-state Potts model is given. The
paper is organized as follows: In section II, by using the
standard representation for the transfer matrices of the $q$-state
Potts model [15] the determinants of the transfer matrices for $n
\times n$  lattices with periodic boundary conditions are
calculated.

\vskip 0.2in \centerline{II. \bf   Determinant of the Potts model
transfer matrix} \vskip 0.1in

 The $q$-state Potts model has served as a valuable model in the
 study of phase transition and critical phenomena. On a lattice,
 or more generally on a graph $G$, at temperature $T$ this model is
 defined by the partition function:
$$
Z(G,q,k)=\sum_{\{\sigma_n\}} e^{-\beta\, H} \eqno(1)$$ \noindent

\noindent  with the Hamiltonian

$$H=-J\sum_{<i,j>} \delta(s_i, s_j) \eqno(2)$$

\noindent  where $\delta(s_i, s_j)$ is the kronecker delta and
$s_i=1,\ldots ,q$ are the spin variables on each vertex $i\in G$
, $\beta=(k_B T)^{-1}$, $k=\beta J $; and $<i,j>$ denotes pairs of
adjacent vertices. Consider an $n \times n$   square lattice with
periodic boundary conditions. For the Ising model, partition
function can be written as product of transfer matrices and the
eigenvalues can be calculated exactly [15,16,17,18,19,20]. There
are also several representations for the transfer matrix of the
$q$-state Potts model [15,21]. Following the method which is used
in [15,16,17,21] for representation of  the transfer matrices of
the two dimensional Ising model and the $q$-state Potts model, we
use a decomposition of the transfer matrices which helps us to
obtain the determinant of the matrices.

\noindent Consider a square lattice of $N=n^2 $ spins consisting
of $ n$ rows and $n$ columns with a toroidal boundary condition.
Let ${\gamma_{\alpha}}\equiv  \{ s_1,s_2, \ldots  , s_n \} $
denote the collection of all spin coordinates of the $\alpha$th
row $(\alpha = 1, \ldots , n)$ with a  toroidal boundary
condition $ {\gamma_{n+1}} \equiv {\gamma_{1}}$. A configuration
of the entire lattice is then specified by $\{ \gamma_1, \ldots ,
\gamma_n\}$. Let $E(\gamma_{\alpha} , \gamma_{\alpha +1} ) $ be
the interaction energy between the $\alpha $th and the $(\alpha
+1 )$th row and $E(\gamma_{\alpha} )$ be the interaction energy
of spins within $\alpha$th row. We can write

$$ E( \gamma , \gamma^\prime )= - J \sum_{k=1}^n \   \delta (s_k , s_k^\prime ) \eqno(3)$$

$$ E(\gamma ) = -J  \sum_{k=1}^n \   \delta (s_k , s_{k+1} ) \eqno(4) $$

\noindent where $ \gamma \equiv \{ s_1, \ldots , s_n \}$ and
$\gamma^\prime \equiv \{ s_1^\prime , \ldots , s_n^\prime \} $
respectively denote the collection of spin coordinates in two
neighboring rows and the partition function is

$$ Z_{PF}= \sum_{\gamma_1} \ldots \sum_{\gamma_n} \exp \ \{ - \beta
\sum_{\alpha =1}^n \ [E(\gamma_\alpha , \gamma_{\alpha +1} ) +
E(\gamma_\alpha ) ] \} \eqno(5) $$

\noindent Let a $q^n \times q^n $ matrix $P$ be so defined that
its matrix elements are

$$ < \gamma \mid P \mid \gamma^{\prime} > \equiv e^{- \beta[E(\gamma , \gamma^\prime ) + E(\gamma )]
} \eqno(6) $$

\noindent Then

$$ Z_{PF}= \sum_{\gamma_1} \ldots \sum_{\gamma_n}\ < \gamma_1 \mid P \mid \gamma_2
> < \gamma_2 \mid P \mid \gamma_3 > \ldots < \gamma_n \mid P \mid \gamma_1
>= {\rm Tr} P^n \eqno(7)$$

\noindent From (3),(4) and (6) we may obtain the matrix elements
of $P $ in the form

$$ < s_1, \ldots , s_n \mid P \mid s_1^\prime, \ldots ,s_n^\prime
> = \prod_{k=1}^n e^{  {k  \delta(s_k, s_{k+1})}}\ e^{  {k
\delta(s_k, s_k^\prime) }} \eqno(8) $$

\noindent Let us define two $q^n\times q^n $ matrices $
V_1^\prime $ and $V_2$ whose matrix elements are given by [15]

$$ <s_1, \ldots ,s_n \mid V_1^\prime \mid s_1^\prime , \ldots ,
s_n^\prime > \equiv \prod _{k=1}^n e^{k \delta (s_k,s_k^\prime)}
\eqno(9) $$

$$ <s_1, \ldots ,s_n \mid V_2 \mid s_1^\prime , \ldots ,
s_n^\prime > \equiv \delta(s_1, s_1^\prime) \ldots
\delta(s_n,s_n^\prime) \prod_{k=1}^n e^{ k \delta (s_k,s_{k+1}) }
\eqno(10) $$

\noindent where $V_2 $ is a diagonal matrix in the present
representation. It is easily verified that $ P=V_2 V_1^\prime $,
or it can be written as

$$< s_1 , \ldots , s_n \mid P \mid s_1^\prime ,\ldots ,s_n^\prime
>=$$
$$
\sum_{s_1^{\prime\prime},\ldots ,s_n^{\prime\prime}} <s_1,\ldots
,s_n \mid V_2 \mid {s_1^{\prime\prime},\ldots
,s_n^{\prime\prime}}> <{s_1^{\prime\prime},\ldots
,s_n^{\prime\prime}} \mid V_1^\prime \mid s_1^\prime , \ldots ,
s_n^\prime > \eqno(11) $$

\noindent Let $A_1 $ and $A_2 $ be two $m \times m$  matrices
whose elements are respectively $<i\mid A_1 \mid j>$ and $<i\mid
A_2 \mid j>$, where $i$ and $j$ independently take on the values
$1,2,..., m$. Then the direct product $ A_1 \otimes A_2 $ is the
$m^2 \times  m^2$ matrix whose matrix elements are

$$ <i_1 i_2  \mid A_1\otimes A_2 \mid
j_1 j_2 >= <i_1\mid A_1 \mid j_1><i_2\mid A_1 \mid j_2>  \eqno(12)
$$

\noindent  This definition can be immediately extended to define
the direct product $A_1\otimes A_2\otimes \cdots \otimes A_n $ of
any number of $m\times m $ matrices $A_1, A_2, \ldots ,A_n$:

$$ <i_1 i_2 \cdots i_n \mid A_1\otimes A_2 \otimes \cdots \otimes A_n \mid
j_1 j_2 \cdots j_n >=$$

$$<i_1\mid A_1 \mid j_1><i_2\mid A_2 \mid j_2> \cdots <i_n\mid A_n \mid j_n> \eqno(13) $$

\noindent  By inspection of (9) it is clear that $V_1^\prime $ is
a product of $n\  q\times q$ identical matrices

$$ V_1^\prime = A\otimes A \otimes \cdots \otimes A \eqno(14)$$

\noindent where

$$ <s\mid A \mid  s^\prime > = e^{k \delta(s,s^\prime)} \eqno(15)$$

\noindent Therefore

$$ A=\pmatrix{{e^k}&{1}&\ldots&{1}\cr
 {1}& e^k & & & \cr
 \vdots& & \ddots& &\cr
 {1}&\ldots& & e^k}=e^k I_{q\times q} + \sigma_{q\times
 q}
\eqno(16) $$

 \noindent where $\sigma$ is a $q\times q$ matrix with zero diagonal
 elements and unit elements on all other entries (note that $ \sigma^2 =(q-2)\  \sigma  + (q-1)\ I $ )
 and $I$ is a
 $q\times q$ unit matrix. Let us represent $A$ by the following
 equation

 $$A = f(k) \ e^{ \tilde {k\over 2} X} \ \ \ \  \  \ \ \ ,  \ \ \  \ \  \ \ \ \
 X^2\equiv I_{q\times q} \eqno(17)$$

\noindent where $ f$ is a function of $k$ and a condition is
imposed on $X$. $\tilde k$ is the dual of $k $ which is given by
the duality relation

$$ e^{- \tilde{k}}={{e^{k}-1}\over {e^{k}+(q-1)}}   \eqno(18)$$

\noindent By considering a linear relation between $X$ and $
\sigma $ ($X_{q\times q }=a \sigma_{q \times q} + b I_{q\times q}
$) we can calculate $f(k) $ and $X$. After a straightforward
calculation we arrive at

$$ X = {2\over q} \sigma + ({2\over q}- 1)I \eqno(19) $$

$$ f(k) = (e^k -1) \ e^{\tilde k \over 2 } \eqno(20)
$$

\noindent Hence

 $$\eqalignno{
 V_{1}^{\prime}&=\big[(e^k -1) \ e^{\tilde k \over 2 }]^{n}
 \,\,\exp{(\ \tilde{k\over 2}\sum_{\alpha=1}^{n}X_{\alpha})}&(21) \cr
 &=\big[(e^k -1) \ e^{\tilde k \over 2 }]^{n}\,V_1\,\, &(22)\cr
 X_{\alpha}&=1\otimes\ldots\otimes1\otimes
 X\otimes1\otimes\ldots\otimes 1
 &(23) \cr}$$
 \noindent  where $X$ is the $\alpha $th factor. In this part we will use the following representation
 for $V_2 $ which is a result of its definition in (10).

$$ V_2 =\prod_{\alpha=1}^n e^{{({k\over q})} \sum_{r=0}^{q-1}  Z_{\alpha}^r Z_{\alpha +1}^{-r}} \eqno(24) $$

$$  Z_\alpha =1\otimes\ldots\otimes1\otimes
 Z\otimes1\otimes\ldots\otimes1 \eqno(25) $$

\noindent where  $Z$ is a diagonal $q\times q$ matrix

$$Z=
\pmatrix{1 & 0 & \ldots & 0 \cr
          0 & w & \ldots & 0 \cr
          \vdots & \vdots & \ddots & \vdots  \cr
          0 & 0 & \ldots & w^{q-1} \cr}\ \ \ \ ,\ \ \ \  Z^q=I_{q\times q}  \ \ \ \ ,
          \ \ \ \ w= e^{{2 \pi i}\over q}   \eqno(26)$$

\noindent   The determinant of the transfer matrix can then be
   calculated using (22) and (24) (note that ${\rm Tr }(A\otimes B) = ({\rm Tr} A )({\rm Tr} B)
   $ and { det}({\rm exp} [A])=exp[{\rm Tr} A])

   $$
   \eqalignno{\det V_1&=\exp\,\bigg[{\rm Tr }
   \big(\tilde{k\over 2}\sum_{\alpha=1}^{n}\,X_\alpha\big)\bigg] &(27)\cr
    &=\exp\bigg[{n \,\tilde{k}\over 2}\,q^{n-1}\,(2-q) \bigg]
   &(28) \cr}$$

\noindent and as $Z_{\alpha}^r $ is traceless for $r\neq 0$ (note
that $\sum_{r=0}^{q-1}w^{(i-j)r} = q \delta_{ij} $)

$$\eqalignno{\det V_2&=\exp\,\bigg[{\rm  Tr}
   \big({k\over q}\sum_{\alpha=1}^{n}\sum_{r=0}^{q-1}\,Z_\alpha^r
   Z_{\alpha+1}^r \big)\bigg] &(29) \cr
 &=\exp\bigg[n\,k\,q^{n-1} \bigg]&(30)\cr
}$$

\noindent
and

$$ \det (V_2 V_{1}^{\prime})=(e^{k}-1)^{n \, q^n}(e^k \, e^{\tilde{k}})^{n
\,q^{n-1}}\eqno(31) $$

\noindent which has already been conjectured in [14]. It may be
interesting to extend these results to lattices with different
boundary conditions. It may also be useful to write a transfer
matrix for the three dimensional Potts model and calculate its
determinant. This decomposition of the transfer matrix may also be
useful for obtaining other exact results for the two dimensional
Potts model and maybe for calculation of its partition function
which is still an unsolved problem.

\vskip 0.4in \centerline{V. \bf   Conclusion} \vskip 0.1in

In this work a conjectured formula by Chang and Shrock [14] for
determinants of the transfer matrices of the q-state Potts model
is proved.

\vskip 0.2in \centerline{\bf \ \ Acknowledgements} \vskip 0.2in

We would like to thank R.J. Baxter for comments and suggestions.
The Authors would like to thank the Isfahan University of
Technology and Institute for Studies in
 Theoretical Physics and Mathematics for the financial support they made available to us.
\vskip 0.2in
\centerline{\bf \ \  References}
\vskip 0.1in

\noindent [1] \ R. B. Potts, Proc. Camb. Phil. Soc. {\bf 48}, 106
(1952).

\noindent [2] \ F. Y. Wu, Rev. Mod. Phys. {\bf 54} (1982) 235.

\noindent [3] \ S. Alexander, Phys. Lett. {\bf A54}, 353 (1975).

\noindent [4] \ A. N. Berker, S. Oslund, and F.Putnam, Phys. Rev.
{\bf B17}, 3650 (1978).

\noindent [5] \ E. Domany et al, Phys. Rev. {\bf B18}, 2209
(1978).

\noindent [6] \ R. J. Baxter, J. Phys. C {\bf 6} L445 (1973).

\noindent [7] \ R. J. Baxter et al, Proc. Roy. Soc. London, Ser.
A {\bf 358}, 535 (1978).

\noindent [8] \ R. J. Baxter, J. Stat. Phys. {\bf 28},1 (1982).

\noindent [9] \ R. J. Baxter, Proc. Roy. Soc. London, Ser. A {\bf
383}, 43 (1982).

\noindent [10] \ M. P. M. den Nijis, J. Phys A {\bf 12}, 1825
(1979); Phys. Rev. {\bf B27}, 1674.

\noindent [11] \ J. L. Black and V. J. Emery, Phys. Rev. {\bf
B23}, 429 (1981).

\noindent [12] \ B. Nienhuis, J. Appl. Phys. {\bf 15}, 199 (1982).

\noindent [13] \ V. S. Dotsenko, Nucl. Phys. {\bf B235}, 54
(1984), and refs therein.

\noindent [14] \ S. Chang and R. Schrok, Physica  A {\bf  296},
234-288 (2001). cond-mat/0011503.

\noindent \ \ \ \ \ \ \ (e.g. Eq. 3.4.39 or 3.70 in Physica A )

\noindent [15] \ R. J. Baxter, Exactly Solved Models in St. Phys.
(Academic Press, 1982).

\noindent [16] \ K. Huang, Statistical Physics, 2nd edition,
(Wiley). Chapter 15.

\noindent [17] \ B. Bergersen and M. Plischke, Equilibrium
Statistical Physics, 2nd edition,

\noindent \ \ \ \ \ \ \ (World Scientific). Chapter 5.

\noindent [18] \  L. Onsager, Phys. Rev. {\bf 65}, 117-49 (1944).

\noindent [19] \  B. Kaufmann, Phys. Rev. {\bf 76}, 1232-1243
(1949).

\noindent [20] \  T. Shultz, D. Mattis  and E. Lieb,  Rev. Mod.
Phys. {\bf 36}, 856 (1964).

\noindent [21] \ P. Martin, Potts Models and related Problems in
St. Phys. (World Scientific).

 \vfill\eject

\bye